\documentclass[prl,twocolumn,showpacs,showkeys]{revtex4}%
\pdfoutput=1
\usepackage{amsfonts}
\usepackage{amsmath}
\usepackage{amssymb}
\usepackage{graphicx}%
\providecommand{\U}[1]{\protect\rule{.1in}{.1in}}
\begin{document}
\preprint{ }
\title[ ]{Leggett-Garg inequalities and the geometry of the cut polytope}
\author{David Avis$^{1,2}$, Patrick Hayden$^1$, and Mark M. Wilde$^1$}
\affiliation{$^1$School of Computer Science, McGill University, Montreal, Quebec, Canada H3A 2A7}
\affiliation{$^2$School of Informatics, Kyoto University, Kyoto, Japan}

\begin{abstract}
The Bell and Leggett-Garg tests offer operational ways to demonstrate that
non-classical behavior manifests itself in quantum systems, and
experimentalists have implemented these protocols to show that classical
worldviews such as local realism and macrorealism are false, respectively.
Previous theoretical research has exposed important connections between more general
Bell inequalities and polyhedral combinatorics. We show here that general Leggett-Garg 
inequalities are closely related to the cut
polytope of the complete graph, a geometric object well-studied in combinatorics. 
Building on that connection, 
we offer a family of Leggett-Garg inequalities that are not
trivial combinations of the most basic Leggett-Garg inequalities. We then show
that violations of macrorealism can occur in surprising ways, by giving an
example of a quantum system that violates the new \textquotedblleft pentagon\textquotedblright\ 
Leggett-Garg inequality but does not violate any of the basic
\textquotedblleft triangle\textquotedblright\ Leggett-Garg inequalities.
\end{abstract}
\volumeyear{2010}
\volumenumber{ }
\issuenumber{ }
\eid{ }
\date{\today}
\startpage{1}
\endpage{ }
\maketitle

Quantum theory offers a radical departure from the classical world, and this
departure manifests itself operationally in the form of Bell~\cite{Be64}\ and
Leggett-Garg inequalities (LGIs)~\cite{LG85}. A Bell inequality bounds the spatial
correlations in any classical theory where observables have definite values
and spacelike separated objects do not influence one another (\textit{local
realism} \cite{epr1935}). An LGI bounds the temporal
correlations in any classical theory in which observables have definite values
and measurement does not disturb the state 
(\textit{macrorealism}). Since these theoretical insights, experimentalists
have observed violations of local realism~\cite{PhysRevLett.49.1804}\ and
macrorealism~\cite{GABLOWP09}\ with quantum optical experiments. Recent theoretical
evidence even suggests that measurement outcomes on biomolecules could violate an LGI~\cite{WMM10}.

The conventional setting for a Bell inequality involves two spacelike
separated parties, say Alice and Bob, each of whom possess a quantum system
$A$ and $B$, respectively. Alice measures one of two dichotomic ($\pm 1$-valued) observables
$A_{1}$ or $A_{2}$ at her end, and Bob measures one of two dichotomic
observables $B_{1}$ or $B_{2}$ at his end. The Clauser-Horne-Shimony-Holt (CHSH)
Bell inequality~\cite{CHSH69} bounds the
following sum of two-point correlation functions in any local realistic
theory:%
\[
\left\langle A_{1}B_{1}\right\rangle +\left\langle A_{1}B_{2}\right\rangle
+\left\langle A_{2}B_{1}\right\rangle -\left\langle A_{2}B_{2}\right\rangle
\leq2.
\]
A bipartite quantum system in an entangled state can violate the above
inequality, demonstrating that the local realistic picture of the universe is false.

Bell inequalities beyond the above conventional two-party, two-observable
setting admit a rich mathematical structure. Peres showed that they correspond
to the facets of a convex polytope, which he called the Bell
polytope~\cite{P99}. It is an example of a correlation polytope,
which have been much studied, see for example Ref.~\cite{pi91} and the 
encyclopedic Ref.~\cite{DL97a}. 
Avis \textit{et al}.~described a relationship between
the Bell polytope and a projecion of the cut\ polytope~\cite{AIIS04,AII06a}, 
a polytope which is isomorphic to the correlation
polytope, and studied in depth in Ref.~\cite{DL97a}.
They were then able to offer 44,368,793 inequivalent tight
Bell inequalities other than those of the CHSH
form for the bipartite setting where each party measures ten
dichotomic observables~\cite{AIIS04}.

The conventional setting for an LGI involves a single
party, say Quinn, who possesses a single quantum system. Quinn measures three
dichotomic observables $Q_{1}$, $Q_{2}$, and $Q_{3}$ as his system evolves in
time \footnote{Equivalently, Quinn could measure rotated observables if his
system does not evolve in time.}. The LGI bounds a 
sum of two-time correlation functions in any macrorealistic theory:%
\begin{equation}
\left\langle Q_{1}Q_{2}\right\rangle +\left\langle Q_{2}Q_{3}\right\rangle
+\left\langle Q_{1}Q_{3}\right\rangle +1\geq0.\label{eq:basic-LG}%
\end{equation}
Quinn can obtain the correlators $\left\langle Q_{1}Q_{2}\right\rangle $,
$\left\langle Q_{2}Q_{3}\right\rangle $, and $\left\langle Q_{1}%
Q_{3}\right\rangle $ with many repetitions of one experiment where he measures
all three observables, or he can obtain them with many repetitions of three
different experiments where each experiment measures only the observables in a
single correlator $\left\langle Q_{i}Q_{j}\right\rangle $.
Note, for example, that if the system
behaves according to the postulates of macrorealism, it should not matter in
which way he obtains the correlators or even if he measures $Q_{2}$ while
obtaining the correlator $\left\langle Q_{1}Q_{3}\right\rangle $. Any quantum
system evolving according to a non-trivial, time-independent Hamiltonian leads
to a violation of the above LGI~\cite{PhysRevLett.101.090403}.

The aim of the present paper is to go beyond the above conventional setting
for a Leggett-Garg experiment and begin exploring the rich mathematical
structure of LGIs with multiple measurements of
dichotomic observables. The inequality in (\ref{eq:basic-LG}) is the simplest
LGI, and it is a \textquotedblleft triangle\textquotedblright \  LGI in the sense that it involves three observables and all
three correlations between them. (This nomenclature will become more clear
later when we explore richer LGIs.) Two works have
already considered multiple measurements in LGIs~\cite{B09,WM10}, but the first work~\cite{B09}\ did not actually generate
any \textquotedblleft new\textquotedblright\ LGIs in the
sense that one can derive all inequalities found there by combining triangle LGIs. The aim of the second work~\cite{WM10}\ was to introduce extra
measurements in order to address the \textquotedblleft clumsiness
loophole\textquotedblright\ in a Leggett-Garg experiment so it did not yield any new inequalities either. We also mention that
another work considered the generalization of LGIs to
higher-dimensional systems~\cite{KB07}, but here we are only concerned with
qubit observables.

In this Letter, we show that strong LGIs are equivalent to facet inequalities
for the cut polytope of combinatorics~\cite{DL97a}. This connection allows us to identify new
classes of LGIs that are not merely combinations of
triangle LGIs. The first interesting LGI that is not a trivial combination of triangle LGIs is a \textquotedblleft pentagon\textquotedblright\  LGI involving the ten different pairwise
correlations between five observables $Q_{1},\ldots,Q_{5}$. We obtain other
non-trivial inequalities for a higher number of observables by exploiting known results
on the facets of the cut polytope. We also propose an experimental setup
including five observables that violates the pentagon LGI,
but in which the pairwise correlations of any three observables do not violate a
triangle LGI. For this example, it is clear that the
standard triangle Leggett-Garg test does not detect the presence
of non-classicality, but a pentagon Leggett-Garg test does indeed detect
non-classicality in the form of a violation.

We structure this work as follows. We first develop the connection between the
LGIs and the cut polytope by reviewing some basic notions
from polyhedral combinatorics. We then discuss our proposed experimental setup
that violates the pentagon LGI but does not violate any
triangle LGI.

\textit{LGIs and the cut polytope}---A macrorealistic
worldview implies that the set of joint probabilities accessible in any
Leggett-Garg experiment involving $n$ observables $Q_{1},\ldots,Q_{n}$ is a
convex polytope. An LGI corresponds to a valid
inequality for this
polytope, namely one that is satisfied by all vectors in the polytope.
The strongest such inequalities are facets of the polytope and
separate macrorealistic from non-macrorealistic
behavior. Facets are those valid inequalities that cannot be obtained
from a positive linear combination of other valid inequalities.
They are the strongest inequalities in the following sense.
A vector violating a (properly normalized)
valid inequality that is not a facet will always
provide a stronger violation of one of the (properly normalized) facets of
which the valid inequality is a positive combination,
hence our interest in finding new facets.
We explain below that the Leggett-Garg polytope for $n$
observables corresponds exactly to the cut polytope for a complete graph with
$n$ nodes.

We begin with some definitions from polyhedral
combinatorics~\cite{DL97a}. Suppose we have an integer $n\geq2$ and a sequence
$(b_{1},b_{2},...,b_{n})$ of integers. Let $b=\sum_{i=1}^{n}b_{i}$,
$k=\sum_{i=1}^{n}|b_{i}|$ and note that $b$ and $k$ have the same parity. We
can define a \textit{k-gonal} inequality over real variables $x_{ij},~1\leq
i<j\leq n$ using these integers:%
\begin{equation}
\sum_{1\leq i<j\leq n}b_{i}b_{j}x_{ij}\leq\left\lfloor\frac{b^{2}}{4}%
\right\rfloor.\label{hyp}%
\end{equation}

Some special classes of the above inequality are of particular interest for us
here. When $b$ is even and some subset of the $b_{i}$ sum to exactly $b/2$,
the corresponding inequality is said to be of \textit{negative type}. When $b$ is odd
and some subset of the $b_{i}$ sum to exactly $\lfloor b/2\rfloor$, the
corresponding inequality is called \textit{hypermetric}. Deza proved that each
$2k$-gonal inequality can be expressed as a positive combination of
$(2k-1)$-gonal inequalities (see, eg. \cite{DL97a}). So in this sense, there
are no \textquotedblleft new,\textquotedblright\ non-redundant inequalities
for even values of $b$.

Now suppose $n=3$. Then the hypermetric inequalities corresponding to the
integer sequences $(1,1,1)$ and $(1,1,-1)$ define, respectively, two types of
triangle inequality:%
\begin{align}
x_{12}+x_{13}+x_{23}  &  \leq2,\label{tri}\\
x_{12}-x_{13}-x_{23}  &  \leq0.
\end{align}
The negative type inequality based on $(1,1,1,1)$ can easily be constructed by
combining four of the first type of triangle inequality.

The set of all triangle inequalities defined on $x_{ij},~1\leq i<j\leq n$
defines a full dimensional polytope, called the \textit{semi-metric polytope}.
The integer solutions to this set of inequalities are all $0/1$-valued, and
are called \textit{cut vectors} because they are equivalent to the edge
incidence vectors of cuts in the complete graph $K_{n}$ \footnote{The complete graph $K_{n}$
is a graph with $n$ vertices and $n(n-1)/2$ edges connecting all vertices to each
other. A (edge) cut of $K_n$ is defined by any subset $S$ of its vertices, and consists of those edges joining a vertex of $S$ to a vertex not in $S$.
The edge incidence vector of a cut is a
binary vector of size $n(n-1)/2$. A component of the vector
is ``zero'' if an edge is not present in the cut, and it is
``one'' otherwise. There are $2^{n-1}$ such vectors.}. The convex hull of
the cut vectors is called the \textit{cut polytope}.
Hypermetric or negative type inequalities for which all $b_{i}=\pm1$ are
called \textit{pure}. It is known that all pure
hypermetric inequalities define facets of the cut polytope~\cite{DL97a}. 

A vertex of the cut polytope represents the correlations
between the values of $n$ variables obtained in a single experiment, where
each value is either 0 or 1.
If there is a joint probability distribution over these $n$ random variables
and the experiment is repeated many times, then the average of the correlations
obtained is a point in the convex hull of these vertices, i.e., a point in
the cut polytope.
Thus,
the facets of the cut polytope collectively describe the correlations that are
accessible in any macrorealistic theory. The triangle inequalities define all
of the facets of the cut polytope for $n=3$ and $n=4$. The first new facet
beyond the triangle inequalities is the pentagon inequality:%
\begin{equation}
\sum_{1\leq i<j\leq5}x_{ij}\leq6. \label{pen}%
\end{equation}

LGIs are typically expressed in terms of the expectations of $\pm1$
random variables, which we denote $Q_{1},Q_{2},...,Q_{n}$. The following
relation allows us to convert the $0/1$ values of the variables $x_{ij}$ to
the $+1$/$-1$ values for the two-time correlations $\left\langle Q_{i}%
Q_{j}\right\rangle $:%
\begin{equation}
x_{ij}=\frac{1-\mathinner{\langle{Q_i Q_j}\rangle}}{2}.\label{rel}%
\end{equation}
We can then convert (\ref{hyp}) to the following inequality:%
\begin{equation}
\sum_{1\leq i<j\leq n}b_{i}b_{j}\mathinner{\langle{Q_i Q_j}\rangle}+\left\lfloor
\frac{\sum_{i=1}^{n}{b_{i}}^{2}}{2}\right\rfloor\geq0.\label{lhyp}%
\end{equation}
Because pure hypermetric inequalities give facets of the cut polytope, the inequality in (\ref{lhyp})
yields facets of the Leggett-Garg polytope when $b_i = \pm 1$.
In this way, the triangle inequalities in (\ref{tri}) become the triangle LGIs \cite{LG85}:%
\begin{align}
\mathinner{\langle{Q_1 Q_2}\rangle}+\mathinner{\langle{Q_1 Q_3}\rangle}+\mathinner{\langle{Q_2 Q_3}\rangle}+1
&  \geq0,\label{ltri}\\
\mathinner{\langle{Q_1 Q_2}\rangle}-\mathinner{\langle{Q_1 Q_3}\rangle}-\mathinner{\langle{Q_2 Q_3}\rangle}+1 \label{ltri2}
&  \geq0.
\end{align}
The second of the
above inequalities is in fact the same as Bell's original inequality
(\cite{Be64}, equation (15)). Since the cut polytope for $n=4$ is completely defined
by triangle inequalities, there are no new strong LGIs
for correlations between 4 random variables.
For $n=5$ however, we obtain the  pentagon LGI 
by rewriting (\ref{pen}) using (\ref{rel}):%
\begin{equation}
\sum_{1\leq i<j\leq5}\mathinner{\langle{Q_i Q_j}\rangle}+2\geq0.\label{lpen}%
\end{equation}
The inequalities (\ref{lhyp}), (\ref{ltri}), (\ref{ltri2}) and (\ref{lpen}), derived from
hypermetric inequalities, define lower bounds on the two-time correlation
functions in any macrorealistic theory.

\textit{Pentagon violation with no triangle violation}---We\ provide a
straightforward experimental setup that violates the pentagon LGI in (\ref{lpen}), but does not violate any of the triangle
LGIs. We assume that the system is noiseless and has
vanishing Hamiltonian so that the dynamics are trivial. We choose as
observables:%
\[
Q_{1}\equiv\sigma_{z},\ \ Q_{2}\equiv\sigma_{\theta},\ \ Q_{3}\equiv\sigma
_{z},\ \ Q_{4}\equiv\sigma_{\theta},\ \ Q_{5}\equiv\sigma_{z},
\]
where $\sigma_{\theta}\equiv\cos\left(  \theta\right)  \sigma_{z}+\sin\left(
\theta\right)  \sigma_{x}$ and $\sigma_{z}$ and $\sigma_{x}$ are
Pauli operators. The inequality in (\ref{lpen}) features ten
two-time correlation functions. As stated before, Quinn can calculate these
correlation functions in one experiment or he can calculate them with ten
different experiments---the way in which he collects the correlation data
should not matter according to the macrorealistic worldview. Also, assuming
macrorealism, Quinn can choose to measure or not measure any of the
observables $Q_{1},\ldots,Q_{5}$ while calculating the correlator
$\left\langle Q_{i}Q_{j}\right\rangle $ because any of these measurements
should not affect the state or its subsequent dynamics according to the
macrorealistic worldview. So, for example, in the calculation of $\left\langle
Q_{1}Q_{5}\right\rangle $, Quinn could measure $Q_{2}$ and this measurement
should not affect the two-time correlation $\left\langle Q_{1}Q_{5}%
\right\rangle $ assuming macrorealism.%
%TCIMACRO{\FRAME{ftbpFU}{3.3909in}{4.9303in}{0pt}{\Qcb{The above figure
%displays the ten experiments that lead to a violation of the pentagon
%LGI. Additionally, any three experiments involving three
%distinct observables do not lead to a violation of the triangle LGI. We depict each two-time correlation function $\left\langle
%Q_{i}Q_{j}\right\rangle $\ to the left of the corresponding experiment and
%display the value of each $\left\langle Q_{i}Q_{j}\right\rangle $ to the right
%of each experiment as a function of the angle $\theta$. Boxed measurements
%indicate that the experimentalist Quinn performs the measurement but that its
%measurement results do not participate in the calculation of the corresponding
%two-time correlation function.}}{\Qlb{fig:leg-exp}}{leggett-experiments.pdf}%
%{\special{ language "Scientific Word";  type "GRAPHIC";
%maintain-aspect-ratio TRUE;  display "USEDEF";  valid_file "F";
%width 3.3909in;  height 4.9303in;  depth 0pt;  original-width 15.0399in;
%original-height 27.2606in;  cropleft "0";  croptop "1";  cropright "1";
%cropbottom "0";  filename 'leggett-experiments.pdf';file-properties "XNPEU";}%
%}}%
%BeginExpansion
\begin{figure}
[ptb]
\begin{center}
\includegraphics[
natheight=27.260599in,
natwidth=15.039900in,
height=4.9303in,
width=3.3909in
]%
{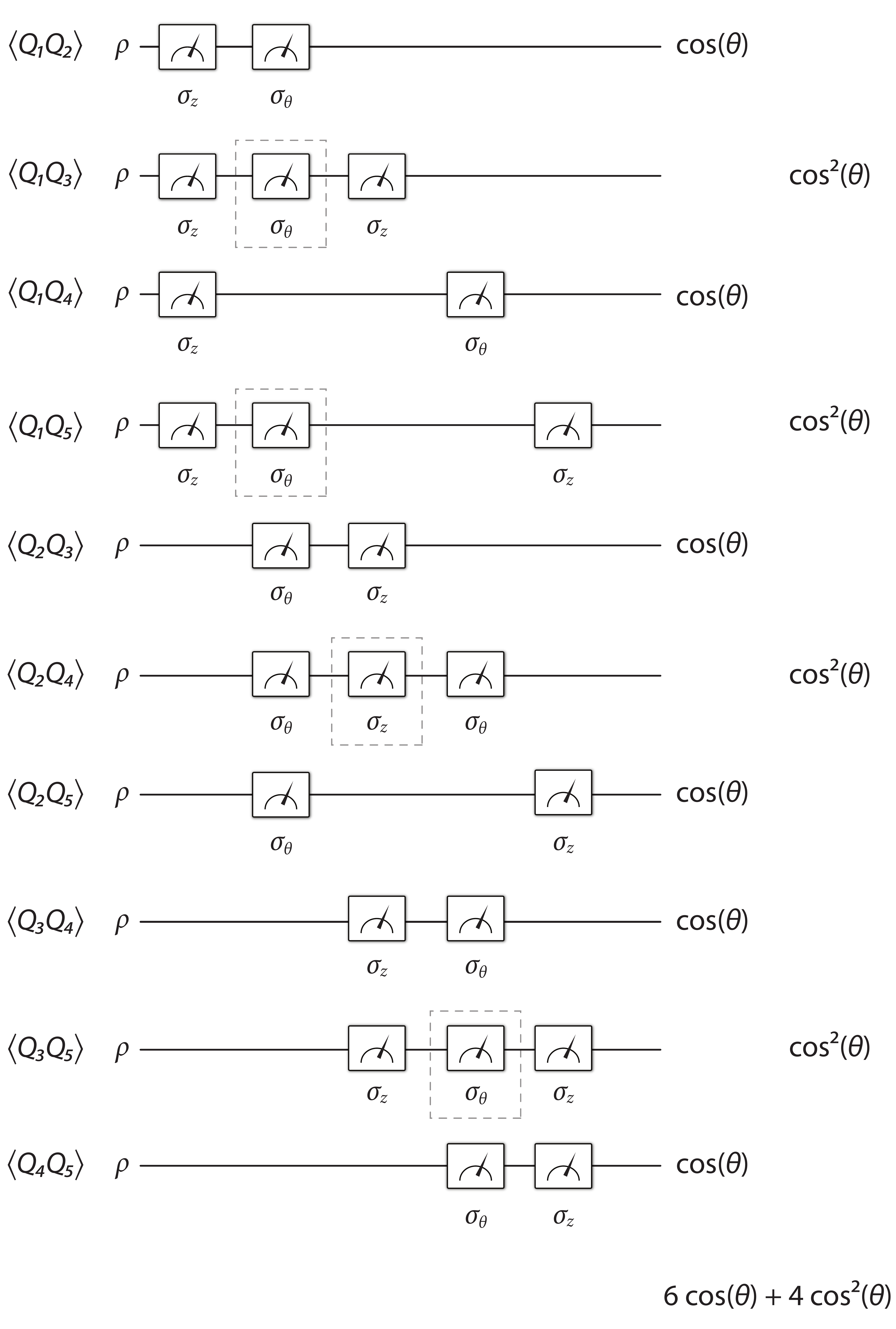}%
\caption{The above figure displays the ten experiments that lead to a
violation of the pentagon LGI. Additionally, any three
experiments involving three distinct observables do not lead to a violation of
the triangle LGI. We depict each two-time correlation
function $\left\langle Q_{i}Q_{j}\right\rangle $\ to the left of the
corresponding experiment and display the value of each $\left\langle
Q_{i}Q_{j}\right\rangle $ to the right of each experiment as a function of the
angle $\theta$. Boxed measurements indicate that the experimentalist Quinn
performs the measurement but that its measurement results do not participate
in the calculation of the corresponding two-time correlation function.}%
\label{fig:leg-exp}%
\end{center}
\end{figure}
%EndExpansion

Figure~\ref{fig:leg-exp}\ displays the ten two-time correlation experiments
that together give a violation of the pentagon LGI. Additionally,
any three of these experiments involving three distinct observables do not
violate the triangle LGI. Below we prove that these
statements hold. We first calculate several relevant quantities. We define the
superoperator $\overline{\Delta}$ as a $\sigma_{z}$ basis dephasing of a qubit
with density operator $\rho$: 
$
\overline{\Delta}\left(  \rho\right)  \equiv\frac{1}{2}\left(  \rho+\sigma
_{z}\rho\sigma_{z}\right)  ,$
and $\overline{\Delta}_{\theta}$ is a $\sigma_{\theta}$ basis dephasing:
$
\overline{\Delta}_{\theta}\left(  \rho\right)  \equiv\frac{1}{2}\left(
\rho+\sigma_{\theta}\rho\sigma_{\theta}\right)  .
$
The following relation is useful: 
$
\overline{\Delta}\left(  \sigma_{\theta}\right)  =\cos\left(  \theta\right)
\sigma_{z},
$
and a similar relation holds by exploiting it: 
$
\overline{\Delta}_{\theta}\left(  \sigma_{z}\right)  =\sigma_{\theta}%
\cos\left(  \theta\right)  .
$
If our input state is the maximally mixed state, then one can check that any
two-time correlation function takes the following form:%
\[
\left\langle Q_{i}Q_{j}\right\rangle =\frac{1}{2}\text{Tr}\left\{
Q_{j}\mathcal{N}\left(  Q_{i}\right)  \right\}  ,
\]
where $\mathcal{N}$ is the map that represents the dynamics between
measurement of $Q_{i}$ and $Q_{j}$. Thus, if there is no measurement between
measurement of $Q_{i}$ and $Q_{j}$, then the two-time correlation function is%
\[
\left\langle Q_{i}Q_{j}\right\rangle =\frac{1}{2}\text{Tr}\left\{  Q_{j}%
Q_{i}\right\}  .
\]
%
% Redundant since \sigma_z = \sigma_0
%
%If there is a measurement of $\sigma_{z}$ between the measurement of $Q_{i}$
%and $Q_{j}$, then%
%\[
%\left\langle Q_{i}Q_{j}\right\rangle =\frac{1}{2}\text{Tr}\left\{  Q_{j}%
%\overline\Delta\left(  Q_{i}\right)  \right\}  .
%\]
If there is a measurement of $\sigma_{\theta}$ between the measurement of
$Q_{i}$ and $Q_{j}$, then%
\[
\left\langle Q_{i}Q_{j}\right\rangle =\frac{1}{2}\text{Tr}\left\{  Q_{j}%
\overline\Delta_{\theta}\left(  Q_{i}\right)  \right\}  .
\]
Using these ideas and the experiments in Figure~\ref{fig:leg-exp}, we
calculate the following two-time correlation functions:%
\begin{eqnarray*}
\left\langle Q_{1}Q_{2}\right\rangle =\left\langle Q_{1}Q_{4}\right\rangle
=\left\langle Q_{2}Q_{3}\right\rangle &=& \cos\left(\theta\right)\\
\left\langle Q_{2}Q_{5}\right\rangle =\left\langle Q_{3}Q_{4}\right\rangle
=\left\langle Q_{4}Q_{5}\right\rangle &=&\cos\left(  \theta\right)  \\
\left\langle Q_{1}Q_{3}\right\rangle =\left\langle Q_{1}Q_{5}\right\rangle
=\left\langle Q_{2}Q_{4}\right\rangle =\left\langle Q_{3}Q_{5}\right\rangle
&=&\cos^{2}\left(  \theta\right)  ,
\end{eqnarray*}
Thus, for these experiments,%
\[
\sum_{1\leq i<j\leq5}\left\langle Q_{i}Q_{j}\right\rangle =6\cos\left(
\theta\right)  +4\cos^{2}\left(  \theta\right)  .
\]
Choosing $\theta$ so that $\cos\left(  \theta\right)  =-3/4$ leads to%
\[
\sum_{1\leq i<j\leq5}\left\langle Q_{i}Q_{j}\right\rangle =-9/4.
\]
This is the smallest that the above \textquotedblleft pentagon
quantity\textquotedblright\ can be for any value of $\theta$, and furthermore
gives a violation of the pentagon LGI. One can also check
that the two-time correlation functions in any three of these experiments
involving three distinct observables never lead to a violation of the standard
LGI because $2\cos\left(  \theta\right)  +\cos^{2}\left(
\theta\right)  \geq-1$ for all $\theta$.

\textit{Conclusion}---The connection between LGIs and the
cut polytope unveils a rich mathematical structure for Leggett-Garg tests with
more than three observables. In particular, (\ref{hyp}) combined with the mapping
(\ref{rel}) gives families of new LGIs that are not
trivial combinations of the original triangle LGIs. Our
example in Figure~\ref{fig:leg-exp}\ shows that it is possible to violate
macrorealism in unexpected ways, e.g., by violating the pentagon LGI without violating any of the triangle LGIs.
Future theoretical work could consider the effects of decoherence on these
violations, similar to the study in Ref.~\cite{WM10}. One could also consider the
maximal violations of LGIs possible in quantum mechanics.
For Bell inequalities, the maximal violations are closely related to the elliptope,
a semidefinite relaxation of the cut polytope defined by the set of negative type inequalities~\cite{AII06a}.
In the case of LGIs, however, an experimentalist is free to perform
a measurement and ignore the outcomes: by exploiting the quantum Zeno effect~\cite{zeno} it then
becomes possible in principle to avoid any meaningful constraint on the correlation functions.
One could nonetheless restrict the experimentalist's actions; perhaps the elliptope
arises from imposing some sensible restrictions.

\acknowledgments

Tobias Fritz recently informed us (private communication)
that he had independently discovered the isomorphism
between LGIs and the cut polytope.
The authors gratefully acknowledge financial support from the Canada
Research Chairs program, the Perimeter Institute, CIFAR, FQRNT's
INTRIQ, MITACS, NSERC, ONR grant N000140811249 and
QuantumWorks. MMW was supported by an MDEIE (Qu\'ebec) PSR-SIIRI international collaboration grant.

\bibliographystyle{unsrt}
\bibliography{Ref}

\begin{thebibliography}{10}

\bibitem{Be64}
John~S. Bell.
\newblock On the {Einstein-Podolsky-Rosen} paradox.
\newblock {\em Physics}, 1:195, 1964.
\newblock Long Island City, New York, USA.

\bibitem{LG85}
Anthony~J. Leggett and Anupam Garg.
\newblock Quantum mechanics versus macroscopic realism: Is the flux there when
  nobody looks?
\newblock {\em Physical Review Letters}, 54(9):857--860, March 1985.

\bibitem{epr1935}
Albert Einstein, Boris Podolsky, and Nathan Rosen.
\newblock Can quantum-mechanical description of physical reality be considered
  complete?
\newblock {\em Physical Review}, 47:777--780, 1935.

\bibitem{PhysRevLett.49.1804}
Alain Aspect, Jean Dalibard, and G\'erard Roger.
\newblock Experimental test of {Bell's} inequalities using time-varying
  analyzers.
\newblock {\em Physical Review Letters}, 49(25):1804--1807, December 1982.

\bibitem{GABLOWP09}
M.~E. Goggin, M.~P. Almeida, M.~Barbieri, B.~P. Lanyon, J.~L. O'Brien, A.~G.
  White, and G.~J. Pryde.
\newblock Violation of the {Leggett-Garg} inequality with weak measurements of
  photons.
\newblock {\em arXiv:0907.1679}, 2009.

\bibitem{WMM10}
Mark~M. Wilde, James~M. McCracken, and Ari Mizel.
\newblock Could light harvesting complexes exhibit non-classical effects at
  room temperature?
\newblock {\em Proceedings of the Royal Society A}, 466(2117):1347--1363, May
  2010.

\bibitem{CHSH69}
John~F. Clauser, Michael~A. Horne, Abner Shimony, and Richard~A. Holt.
\newblock Proposed experiment to test local hidden-variable theories.
\newblock {\em Physical Review Letters}, 23(15):880--884, October 1969.

\bibitem{P99}
Asher Peres.
\newblock All the {Bell} inequalities.
\newblock {\em Foundations of Physics}, 29(4):589--614, April 1999.

\bibitem{pi91}
Itamar Pitowski.
\newblock Correlation polytopes: Their geometry and complexity.
\newblock {\em Mathematical Programming}, 50(3):395--414, 1991.

\bibitem{DL97a}
Michel~Marie Deza and Monique Laurent.
\newblock {\em Geometry of Cuts and Metrics}.
\newblock Springer, May 1997.

\bibitem{AIIS04}
David Avis, Hiroshi Imai, Tsuyoshi Ito, and Yuuya Sasaki.
\newblock Two-party {Bell} inequalities derived from combinatorics via
  triangular elimination.
\newblock {\em Journal of Physics A: Mathematical and General}, 38(50):10971,
  2005.

\bibitem{AII06a}
David Avis, Hiroshi Imai, and Tsuyoshi Ito.
\newblock On the relationship between convex bodies related to correlation
  experiments with dichotomic observables.
\newblock {\em Journal of Physics A: Mathematical and General}, 39(36):11283,
  2006.

\bibitem{PhysRevLett.101.090403}
Johannes Kofler and \ifmmode \check{C}\else~\v{C}\fi{}aslav Brukner.
\newblock Conditions for quantum violation of macroscopic realism.
\newblock {\em Physical Review Letters}, 101(9):090403, August 2008.

\bibitem{B09}
Marco Barbieri.
\newblock Multiple-measurement {Leggett-Garg} inequalities.
\newblock {\em Physical Review A}, 80(3):034102, September 2009.

\bibitem{WM10}
Mark~M. Wilde and Ari Mizel.
\newblock Addressing the clumsiness loophole in a {Leggett-Garg} test of
  macrorealism.
\newblock {\em arXiv:1001.1777}, 2010.

\bibitem{KB07}
Johannes Kofler and \ifmmode \check{C}\else~\v{C}\fi{}aslav Brukner.
\newblock Classical world arising out of quantum physics under the restriction
  of coarse-grained measurements.
\newblock {\em Physical Review Letters}, 99(18):180403, November 2007.

\bibitem{zeno}
E.~C.~G. Sudarshan and B.~Misra.
\newblock The {Z}eno's paradox in quantum theory.
\newblock {\em Journal of Mathematical Physics}, 18(4), 1977.

\end{thebibliography}

\end{document}